\documentclass[floatfix,aps,prd,twocolumn,groupedaddress,showpacs,showkeys,superscriptaddress,preprintnumbers]{revtex4-1}
\usepackage{graphicx}
\usepackage{bm}
\usepackage{amssymb,amsmath}
\usepackage{epsfig}

\def\beqn{\begin{eqnarray}}
\def\eeqn{\end{eqnarray}}
\def\barr{\begin{array}}
\def\earr{\end{array}}
\def\btab{\begin{tabular}}
\def\etab{\end{tabular}}
\def\bite{\begin{itemize}}
\def\eite{\end{itemize}}
\def\bcen{\begin{center}}
\def\ecen{\end{center}}

\def\eq{\begin{equation}}
\def\ee{\end{equation}}
\def\nn{\nonumber}

\def\q2dagger{q_2\hspace{-0.35cm}/\;}

\begin{document}

\title{Bounds on rare decays of $\eta$ and $\eta^\prime$ mesons from the neutron EDM}
\author{Alexey S.~Zhevlakov} 
\email{zhevlakov@phys.tsu.ru}
\affiliation{Department of Physics, Tomsk State University, 634050 Tomsk, Russia} 
\affiliation{Matrosov Institute for System Dynamics and 
Control Theory SB RAS Lermontov str., 134, 664033, Irkutsk, Russia } 
\date{\today}

\author{Mikhail Gorchtein} 
\email{gorshtey@kph.uni-mainz.de}
\affiliation{Institut f\"ur Kernphysik, Johannes Gutenberg-Universit\"at, Mainz, Germany 
and PRISMA Cluster of Excellence, Johannes Gutenberg-Universit\"at, Mainz, Germany} 
\author{Astrid~N.~Hiller~Blin} 
\email{astridblin.4@gmail.com} 
\affiliation{Institut f\"ur Kernphysik, Johannes Gutenberg-Universit\"at, Mainz, Germany 
and PRISMA Cluster of Excellence, Johannes Gutenberg-Universit\"at, Mainz, Germany}

\author{Thomas Gutsche}
\email{thomas.gutsche@uni-tuebingen.de}
\affiliation{Institut f\"ur Theoretische Physik,
Universit\"at T\"ubingen,
Kepler Center for Astro and Particle Physics,
Auf der Morgenstelle 14, D-72076 T\"ubingen, Germany}

\author{Valery E. Lyubovitskij} 
\email{valeri.lyubovitskij@uni-tuebingen.de}
\affiliation{Department of Physics, Tomsk State University, 634050 Tomsk, Russia} 
\affiliation{Institut f\"ur Theoretische Physik,
Universit\"at T\"ubingen,
Kepler Center for Astro and Particle Physics,
Auf der Morgenstelle 14, D-72076 T\"ubingen, Germany}
\affiliation{Departamento de F\'\i sica y Centro Cient\'\i fico
Tecnol\'ogico de Valpara\'\i so-CCTVal, Universidad T\'ecnica
Federico Santa Mar\'\i a, Casilla 110-V, Valpara\'\i so, Chile}
\affiliation{Laboratory of Particle Physics, Tomsk Polytechnic University,
634050 Tomsk, Russia}

\date{\today}

\pacs{12.39.Fe, 13.25.Jx, 14.40.Be, 14.65.Bt}
\keywords{CP-violation, electric dipole moment, nucleons, mesons}

\begin{abstract} 

We provide model-independent bounds on the rates of rare decays 
$\eta (\eta^\prime) \to \pi\pi$ based on experimental limits on 
the neutron electric dipole moment (nEDM). 
Starting from phenomenological $\eta(\eta^\prime) \pi\pi$ couplings, 
the nEDM arises at two loop level. 
The leading-order relativistic ChPT calculation with the minimal 
photon coupling to charged pions and a proton inside 
the loops leads to a finite, counter term-free result. 
This is an improvement upon previous estimates which 
used approximations in evaluating the two loop contribution and 
were plagued by divergences. While constraints on the 
$\eta(\eta^\prime) \pi\pi$ couplings in our phenomenological approach 
are somewhat milder than in the picture with the QCD $\theta$-term, 
our calculation means that whatever the origin of these couplings, 
The decays $\eta (\eta') \to 2 \pi$ will remain unobservable in the near future.

\end{abstract}

\maketitle

\section{Introduction}
The observed matter-antimatter asymmetry in the universe indicates that at some 
early stage in the evolution of the universe  the $CP$-symmetry, 
an exact balance of the rates for processes that involve particles and antiparticles, 
should have been broken \cite{Sakharov:1967dj}. 
However, until the discovery of $CP$-violation in $K$-meson decays, the $CP$-symmetry 
was believed to be an exact symmetry of the Standard Model (SM). 
The explanation for this $CP$-violation problem was found in the 
electroweak sector, involving $CP$-violating (CPV) phases of 
the Cabibbo-Kobayashi-Maskawa (CKM) 
quark-mixing matrix which allowed to accommodate the observations. 
Apart from meson decays, CPV interactions would also induce 
a static electric dipole moment for a particle with  spin. 
Presently, there is a large number of experiments performing 
precise measurements of EDMs of 
hadrons, nuclei, atoms, and molecules~\cite{Chupp:2017rkp}. 
Furthermore, there are also data on CPV meson decays 
(see, e.g., Ref.~\cite{Aaij:2016jaa}). 

In the SM, the EDM may arise due to the CPV phases of the CKM matrix. 
The latest SM prediction for the nEDM~\cite{Seng:2014lea} 
is $|d_n^{\rm CKM}|\approx (1 - 6) \times 10^{-32}\, \text{e}\cdot \text{cm}$. 
This range corresponds to uncertainties of low-energy constants involved 
in the calculations based on the heavy-baryon effective Lagrangian. 
Apart from the CPV in the quark mixing, the SM has no dynamical source of CP violation. 
Current experiments aiming at measuring  the neutron and lepton EDMs 
are sensitive to a signal which is several orders of magnitude larger than that 
allowed in the SM~\cite{Afach:2015sja,PDG}
\begin{align}
\label{dnE_data}
|d^E_n|<2.9\times 10^{-26}\,e\cdot{\rm cm}.
\end{align}
An observation of a non-zero EDM in the near future 
would thus point to a non-SM origin of CP violation.

In the strong-interaction sector, the nEDM is induced by 
the CPV $\theta$-term of quantum chromodynamics (QCD)
\begin{align}
\Delta\mathcal{L} = \theta \frac{g_s^2}{32\pi^2}G^a_{\mu\nu}\tilde{G}^{a\mu\nu},
\end{align}
where $g_s$ is the QCD coupling constant, and $G_{\mu\nu}^a$ and  
$\tilde{G}^{a\mu\nu} = \frac{1}{2}\epsilon^{\mu\nu\alpha\beta} 
G^{a}_{\alpha\beta}$ are the usual stress tensor of the 
gluon field and its dual.  The $\theta$-term preserves the renormalizability 
and gauge invariance of QCD, 
but breaks the P- and T-parity invariance. It plays an important role in QCD, 
e.g., for the QCD vacuum, the topological charge, and the solution of the 
$U(1)_A$ problem of the mass of the $\eta^{\prime}$ meson~(see, e.g., 
Refs.~\cite{Diakonov:1981nv,Witten:1979vv}). 
An explanation to the apparent smallness of the $\theta$ coupling (solution for
the strong CP-violation problem) was proposed by Peccei and Quinn~\cite{Peccei:1977hh}. 
They suggested $\theta$ to be a field $\theta(x)$, and decomposed it into 
an axial field $a(x)$ (axion) that preserves $CP$ conservation, and a small constant 
$\bar\theta$ that encodes the CPV effect. 
For a recent overview see, e.g., Ref.~\cite{Castillo-Felisola:2015ema}. 

The non-zero $\bar\theta$ generates a number of hadronic CPV interactions, 
e.g., a CPV $\pi NN$ coupling. At one loop, this coupling leads to a non-zero EDM 
of the nucleon. The early calculation of Ref.~\cite{Crewther:1979pi} led to 
a constraint on $\bar\theta\lesssim6\times10^{-10}$ based on the experimental bounds 
on the nEDM existing at that time. Other examples of CPV interactions among hadrons 
that arise in presence of the non-zero $\bar\theta$ are the decays 
$\eta(\eta') \to 2\pi$. In the picture where the entire  CP violation 
in hadronic interactions is due to the $\bar\theta$ term, the corresponding 
branching ratio of the order of $\sim10^{-17}$ is unobservable. 
Recent advances in experimental techniques and the possibility to produce large numbers 
of $\eta$ and $\eta'$ mesons at MAMI (Mainz) and 
Jefferson Lab (Newport News)~\cite{Ghoul:2015ifw} has led experimentalists to look for 
or at least set more stringent constraints on rare decays of $\eta$ and 
$\eta'$ mesons. 
Current experimental upper limits read~\cite{PDG,Aaij:2016jaa}
\begin{align}
\rm{Br}(\eta\to\pi\pi) &<
\left\{
\begin{array}{c}
1.3\times10^{-5},\;\pi^+\pi^-\\
3.5\times10^{-4},\;\pi^0\pi^0
\end{array}
\right. \nonumber\\
&\left . \right. \label{eq:explimits}\\
\rm{Br}(\eta'\to\pi\pi)  &<
\left\{
\begin{array}{c}
1.8\times10^{-5},\;\pi^+\pi^-\\
4\times10^{-4},\;\pi^0\pi^0
\end{array}
\right. \nonumber 
\end{align}
These bounds indicate that any signal observed within the $\sim13-14$ orders of magnitude between 
the existing experimental bounds and the strong CPV expectations could be interpreted 
as an unambiguous signal of New Physics. 

Given the experimental constraints 
on the nEDM it is also informative to ask how large a CPV $\eta^{(')}\pi\pi$ 
interaction generated by an unspecified New Physics mechanism could be. 
This question was raised in Ref.~\cite{Gorchtein:2008pe} and revisited 
in Ref.~\cite{Gutsche:2016jap}. 
In those works, effective CPV $\eta^{(')} NN$ couplings 
were generated from an effective CPV 
$\eta^{(')}\to \pi\pi$ coupling via a pion loop. 
At a second step, those CPV couplings were 
used to generate the nEDM, again at one-loop. 
Because both the neutron and the $\eta$'s 
have no charge, the only way to couple an external photon to obtain the nEDM was 
magnetically. As a result, each loop is logarithmically divergent, leading to the 
need of counterterms which made the results less conclusive. 

In this paper, we opt for a direct two-loop calculation and account for the minimal photon 
coupling to charged particles inside the loop. We use the leading-order ChPT Lagrangian for 
the coupling between pseudoscalar mesons and nucleons. CP violation is assumed 
to stem solely from the $\eta^{(\prime)}\pi\pi$ coupling. 
Anticipating the findings of our work, we obtain a resulting contribution   
to the nEDM which is UV-finite. 
We are thus able to derive very robust constraints on 
the CPV $\eta^{(\prime)}\to \pi\pi$ decay 
branching ratios from the tight experimental bounds on the nEDM,
\begin{align}
&\mathrm{Br}(\eta \to \pi^+\pi^-)   < 5.3 \times 10^{-17} \,,\nonumber\\
&\mathrm{Br}(\eta \to \pi^0\pi^0)   < 2.7 \times 10^{-17} \,,\nonumber\\
&\label{G_etapp_intro} \\
&\mathrm{Br}(\eta'\to \pi^+\pi^-)   < 5.0 \times 10^{-19} \,,\nonumber\\
&\mathrm{Br}(\eta'\to \pi^0\pi^0)   < 2.5 \times 10^{-19} \,. \nonumber
\end{align}
It makes the observation of these decay channels hardly possible, 
independent of the particular mechanism that may lead to the generation of such 
an interaction. 
While previous calculations~\cite{Gorchtein:2008pe,Gutsche:2016jap} 
contained an uncertainty due to the divergences in chiral loops, 
this work represents an exact LO chiral result. As compared to the QCD 
$\theta$-term constraints on $\eta^{(\prime)}\to \pi\pi$ decays, the above 
bounds are only slightly less stringent. 

The paper is organized as follows. In Sec.~II we discuss the CPV couplings 
of the $\eta$ and $\eta^\prime$ mesons with two pions. 
In Sec.~III we present the calculation of the nEDM at two loops with leading 
order ChPT meson-nucleon interaction, 
and refer details of the two-loop calculation to Appendix~\ref{appendixA}. 
In Sec.~IV we derive the upper bounds for the $\eta\to 2\pi$ and 
$\eta^\prime \to 2\pi$ decay rates. 

\section{CPV decay constants}   

We begin by considering the rare CPV decays $\eta(\eta') \to 2\pi$. 
For the masses and full widths of the $\eta$ and $\eta'$ mesons we 
use the PDG values~\cite{PDG}: 
$m_\eta=547.862\pm0.017$ MeV, $\Gamma_\eta^\text{full}=1.31\pm0.05$ keV and
$m_{\eta'}=957.78\pm0.06$ MeV, $\Gamma_{\eta'}^\text{full}=0.196\pm0.009$ MeV.  

\begin{figure}[h]
\includegraphics[scale=.75]{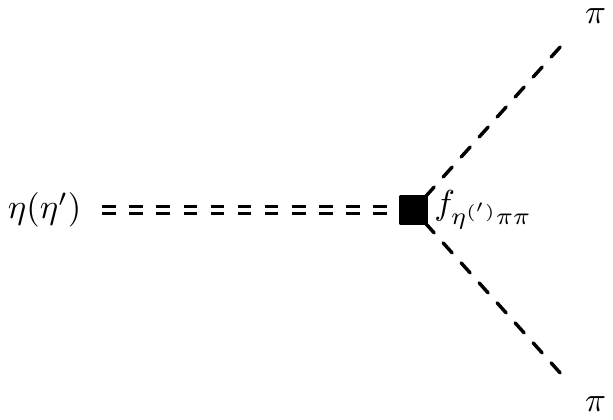}	
\vspace*{-17cm}
\caption{The $\eta(\eta')\to\pi\pi$ decay process. 
The solid square represents the CPV vertex.}
\label{fig:etapipi}
\end{figure}

The effective Lagrangian that generates the P- and T-violating processes 
$\eta(\eta') \to 2\pi$ (see Fig.~\ref{fig:etapipi}) has the form
\beqn
{\cal{L}}=f_{H\pi\pi} m_H H \vec{\pi\,}^2\,,  \quad H = \eta, \eta'\,, 
\label{etapipiL}
\eeqn
\noindent
where $m_H$ is the mass of the $\eta(\eta')$ meson, the pion field $\vec\pi$ 
is a isovector, $f_{H\pi\pi}$ is the corresponding coupling constant 
chosen to be dimensionless and defined for pions and $\eta(\eta')$ mesons on 
the mass shell. The values of the $f_{\eta(\eta')\pi\pi}$ are related to the 
corresponding branching ratios according to
\beqn
{\rm Br}(H \to \pi\pi)=\frac{\Gamma_{H\to\pi\pi}}{\Gamma_H^{full}}
= n_\Gamma \, \frac{\sqrt{m_H^2-4m_\pi^2}}{4\pi\Gamma_H^{full}}f_{H\pi\pi}^2\,.
\label{eq:etapipi_decay_coupling}
\eeqn 
The factor $n_\Gamma$ is $1/2$ for the $\pi^0\pi^0$ and 1 for the $\pi^+\pi^-$ channel 
and reflects the Bose statistics for identical particles in the final state.

There are two possible generic mechanisms for the generation of these effective 
Lagrangians. The first scenario, which is fully explored in literature, 
is the solution to the strong $U(1)_A$ problem in terms of the QCD $\theta$-term 
that generates both the $\eta\to\pi\pi$ decay and the nucleon 
EDM~\cite{Crewther:1979pi,Pich:1991fq,Shifman:1979if}. 
The effective $\eta^{(')}\pi\pi$ couplings in this scenario are given 
by~\cite{Crewther:1979pi,Shifman:1979if}
\beqn
f_{\eta\pi\pi}^\theta&=&-\frac{1}{\sqrt{3}} \, 
\frac{\theta\, m_\pi^2\, R}{F_\pi\, m_\eta \, (1+R)^2}\,,\\
f_{\eta'\pi\pi}^\theta&=&-\sqrt{\frac{2}{3}} \, 
\frac{\theta\, m_\pi^2\, R}{F_\pi \, m_{\eta'} \, (1+R)^2}\,,\label{eq:f-theta}
\eeqn
where $\theta$ is the QCD vacuum angle, $R = m_u/m_d$ is the 
ratio of the $u$ and $d$ current quark masses, 
$F_\pi=92.4$ MeV is the pion decay constant, and $m_\pi=139.57$ MeV is 
the charged pion mass. In this scenario, the decay constant is proportional 
to the $\theta$-term, which is tightly constrained by the experimental 
bounds on the neutron EDM~\cite{Harris:1999jx,Baker:2006ts}. 
It is also seen that $\eta^{(')}\pi\pi$ couplings vanish in the chiral limit 
$m_\pi\to0$ resulting in an additional suppression. 
As a result, the bound for the decay constants in Eq.~(\ref{etapipiL}) is 
$(f_{\eta\pi\pi},f_{\eta^{\prime}\pi\pi})\sim(0.03\,\theta,0.05\,\theta)$. 
The EDM bound 
$\theta < 6\cdot 10^{-10}$~\cite{Crewther:1979pi,Pospelov:1999ha,Balla:1999vx} 
makes experimental searches for the $\eta(\eta^\prime)$ decaying into two pions 
hopeless. 

The second scenario corresponds to the situation where the EDM and the CPV 
$\eta\to\pi\pi$ vertices are generated by two distinct mechanisms, without specifying 
details of a particular model in which this scenario would be realized.
Given the interest in addressing these decay channels experimentally 
at Jefferson Lab~\cite{Ghoul:2015ifw}, it is informative to inquire, how much room there is 
for New Physics contributions that could lead to anomalously large $\eta\pi\pi$ coupling 
constants. The unknown New Physics mechanism would then 
generate a non-zero $f_{\eta\pi\pi}$, which through pseudoscalar meson couplings 
to the nucleon generates the EDM at the two-loop level. 

\section{Neutron EDM induced by CPV couplings} 

The  electromagnetic  nucleon vertex  in presence of $CP$-violation 
is written in terms of Dirac, Pauli and electric dipole form factors 
$F_E(Q^2), F_M(Q^2), F_D(Q^2)$,  
respectively,
\begin{align}
&\bar{u}_N(p_2)\Gamma^\mu(p_1,p_2)u_N(p_1)=
\bar{u}_N(p_2)\left[\gamma^{\mu}F_E(Q^2)  \right.\nonumber \\ &
+\,\left.\frac{i\sigma^{\mu\nu}k_{\nu}}{2m}F_M(Q^2)
+\frac{i\sigma^{\mu\nu}k_{\nu}\gamma_5}{2m_N}F_D(Q^2)\right]u_N(p_1).
\label{vertex}
\end{align} 
Here, $Q^2=-k^2=-(p_2-p_1)^2$, $m_N$ is the nucleon mass, 
$\gamma^\mu$, $\gamma_5$ are the Dirac matrices, and
$\sigma^{\mu\nu} = \frac{i}{2}[\gamma^\mu,\gamma^\nu]$. 
The electric dipole moment of the neutron is defined as $d^E_n=-F_D(0)/(2m_N)$. 

For calculating the pseudoscalar meson loops we use the non-derivative 
pseudoscalar (PS) couplings between mesons and nucleons. 
The pseudoscalar approach is obtained from the more commonly used pseudovector (PV) 
theory by means of a well-known chiral rotation of the nucleon fields. 
The two theories are equivalent 
(see details in Refs.~\cite{Lyubovitskij:2001fv,Lyubovitskij:2001zn,Lensky:2009uv}), 
and at leading order the only term in the Lagrangian involving pion and nucleon fields 
is
\beqn
{\cal L}_{\pi NN}^{\rm PS} &=& 
g_{\pi NN} \, \bar{N} \, i\gamma_5 \vec{\pi} \vec{\tau} \, N\,, \quad 
g_{\pi NN} = \frac{g_A}{F_\pi} m_N,\nn\\
{\cal L}_{\eta NN}^{\rm PS} &=& 
g_{\eta NN} \, \bar{N} \, i\gamma_5 \eta \, N\,,
\label{const_Vert}
\eeqn
where $g_A \approx 1.275$ is the nucleon axial charge and $F_\pi\approx92.4$ MeV 
the pion decay constant. In $SU(3)$  limits, 
the $\eta N$ couplings $g_{\eta^{(\prime)} NN}$ 
would also be related to the respective axial couplings $g_A^\eta$ and 
decay constants $F_\eta$, and similarly for $\eta'$. 
However, $SU(3)$ symmetry appears to be significantly broken for  
these couplings
and recent analyses of $\eta$ and $\eta'$ photoproduction on nucleons
suggests much smaller values~\cite{Tiator:2018heh}:
\beqn 
g_{\eta NN}\approx  g_{\eta' NN}\approx0.9\,. 
\eeqn 

Using the ingredients specified above, we  can calculate the induced nEDM. 
The advantage of the pseudoscalar as compared to the pseudovector pion-nucleon theory 
is two-fold. Firstly, because the coupling is non-derivative the result is finite. 
Secondly, the number of graphs to be calculated at leading order is reduced 
significantly because the only way to couple the electromagnetic field to the 
pion field is minimally to the charged pion lines inside the loop.
Unlike in the PV theory where the contact (Kroll-Ruderman) $\gamma\pi NN$
interaction term appears in the leading order chiral Lagrangian, in  
the pseudoscalar theory
this term is generated at the level of matrix element at order $1/F_\pi$
and the same is true for the $\gamma\pi\pi NN$ term appeared
at order $1/F_\pi^2$~\cite{Lyubovitskij:2001fv,Lyubovitskij:2001zn}.

The full set of two-loop Feynman diagrams to be calculated is shown 
in Fig.~\ref{two_loop}. 
Only diagrams that contribute to the nEDM are displayed. For instance, 
the class of diagrams that involve the contact $\pi\pi NN$ coupling 
gives no contribution to the nEDM and is dropped from Fig.~\ref{two_loop}. 
Among those diagrams that contribute there are further symmetry considerations 
that allow to reduce the number of independent graphs. 
From hermitian conjugation of matrix elements, 
after using replacements of nucleon momenta $p_1 \leftrightarrow p_2$ and  the inverse of the
photon momentum $k \leftrightarrow -k$ we show result in the following relations: 
\beqn 
& &d_n^{E;a} = d_n^{E;b}\,, \quad 
   d_n^{E;c} = d_n^{E;d}\,, \quad 
   d_n^{E;e} = d_n^{E;l}\,, \nonumber\\
& &d_n^{E;f} = d_n^{E;k}\,, \quad 
   d_n^{E;g} = d_n^{E;h}\,. 
\eeqn 
Therefore, the total contribution to the nEDM is 
\beqn 
d_n^E = 2 \, (d_n^{E;a} + d_n^{E;c} + d_n^{E;e} + d_n^{E;f} + d_n^{E;g}) \,.
\eeqn 
The detailed calculation of the two-loop diagrams is reported in Appendix~\ref{appendixA}. 
\begin{widetext}

	\begingroup
	\squeezetable
\begin{figure}		
\includegraphics[width=1.05\textwidth, trim={2cm 21,5cm 0cm 1cm},clip]{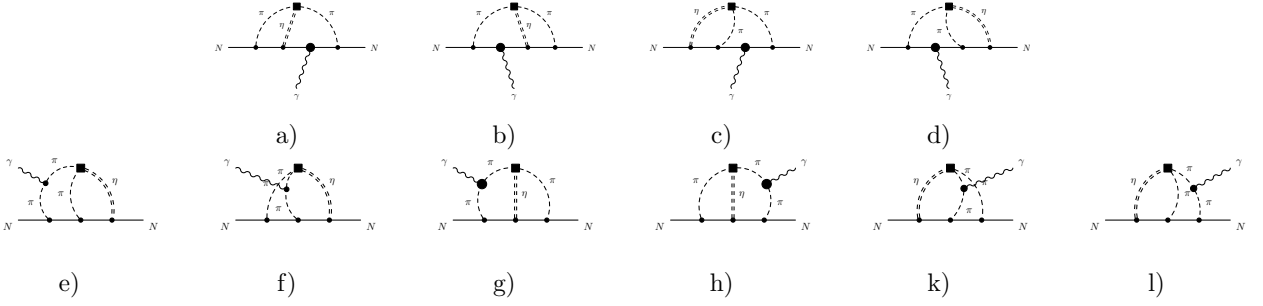}
\caption{Diagrams describing the nEDM. 
The interaction with the external electromagnetic field occurs through 
the minimal electric coupling to charged baryon or meson fields.  
The solid square denotes the CPV $\eta\,\pi^+\pi^-$ vertex.}\label{two_loop}
\end{figure}
	\endgroup	
\end{widetext}

\section{Results and discussion} 
Upon evaluating the two-loop diagrams we obtain an expression for the nEDM induced by 
the CPV $\eta(\eta')\pi\pi$ couplings 
via meson loops and minimal coupling to the electromagnetic field, 
\begin{align}
d_n^E &\simeq \frac{e g_{\pi NN}^2}{(4\pi)^4}\left[9.3 \, 
\frac{ g_{\eta NN}m_\eta}{m_N^2} f_{\eta\pi\pi}+ 6.4\,  
\frac{g_{\eta^\prime NN} m_\eta^\prime}{m_N^2} f_{\eta^\prime\pi\pi}\right] 
\nonumber\\
&\simeq(c_\eta f_{\eta\pi\pi} + 
c_{\eta'} f_{\eta^\prime\pi\pi})\times 10^{-16} \text{e} \cdot \text{cm}\,,
\nonumber\\
c_\eta &= 6.7\,, \quad c_{\eta'} = 7.9 \,. 
\end{align}
The numerical difference between 
the two coefficients arises due to the $\eta$ and $\eta^\prime$ mass 
dependence in the loop integrals. Note that the two coefficients do not contain 
chiral divergences $\sim1/m_\pi$ or $\sim\ln m_\pi$, and are evaluated by setting 
the pion mass to zero. For completeness, 
in Table~\ref{tab:ceta_cetap} we present the partial 
contributions of the diagrams to the couplings $c_\eta$ and $c_{\eta'}$. 

\begin{table}[htb]
\begin{center}
\caption{Contributions of diagrams to $c_\eta$ and $c_{\eta'}$.} 

\vspace*{.15cm}

\def\arraystretch{1.25}
\begin{tabular}{c|c|c}
\hline
Diagram & $c_{\eta}^a$  
        & $c_{\eta'}^a$ \\
\hline
  a(b)  & 0.58 & 0.71 \\
  c(d)  & 0.56  & 0.67  \\
  e(l)  & 1.1  & 1.3  \\
  g(h)  & 1.0  & 1.1  \\
  f(k)  & 0.1  & 0.1  \\
\hline 
Total   & 6.7  & 7.9  \\
\hline
\end{tabular}
\label{tab:ceta_cetap}
\end{center}
\end{table}

Using now the current experimental bound $|d^E_n|<2.9\times 10^{-26}\,e\cdot{\rm cm}$ 
and assuming that the $\eta$ and $\eta'$ couplings to two pions are independent, 
we deduce  upper bounds on the coupling constants, 
\beqn
&|f_{\eta\pi\pi}(m_\eta^2)|           < 4.3 \times 10^{-11},\\
&|f_{\eta^\prime\pi\pi}(m_{\eta'}^2)| < 3.7 \times 10^{-11}\,.
\label{f_etapp}
\eeqn

These translate into  upper bounds for the respective branching ratios, 
\begin{align}
&\mathrm{Br}(\eta \to \pi^+\pi^-)   < 5.3 \times 10^{-17} \,,\nonumber\\
&\mathrm{Br}(\eta \to \pi^0\pi^0)   < 2.7 \times 10^{-17} \,,\nonumber\\
&\label{G_etapp} \\
&\mathrm{Br}(\eta'\to \pi^+\pi^-)   < 5.0 \times 10^{-19} \,,\nonumber\\
&\mathrm{Br}(\eta'\to \pi^0\pi^0)   < 2.5 \times 10^{-19} \,, \nonumber
\end{align}
which strongly are reduced in comparison to existing experimental limits of Eq.~(\ref{eq:explimits}).
While future and ongoing measurements of the rare decay widths of the $\eta$ 
and $\eta^\prime$ into pion pairs may improve the limits  of Eq.~(\ref{eq:explimits}), 
our results show that no finite signal of CP violation in these processes should be expected 
at the currently accessible level of precision.  
A similar conclusion can be made about the decays of 
the $\eta$ and $\eta^\prime$ into four pions~\cite{Guo:2011ir}. 

If we compare the values obtained for $f_{\eta\pi\pi}$ and 
$f_{\eta^\prime\pi\pi}$ with Eq.~(\ref{eq:f-theta}), 
we can deduce an upper limit for the $\bar\theta$ parameter in the 
Peccei-Quinn mechanism, 
\begin{align}
\label{set1}
\bar\theta^{\eta}  <  8.4 \cdot 10^{-10} \,,\quad 
\bar\theta^{\eta'} <  9.0 \cdot 10^{-10} \,.
\end{align} 
Here we use  the ratio $R = m_u/m_d = 0.556$ 
of the canonical set of the quark masses 
in ChPT~\cite{Gasser:1982ap}: 
$m_u = 5$ MeV, $m_d = 9$ MeV~\cite{Gasser:1982ap} at scale 1 GeV. 
The limit 
\begin{align}
\label{set2}
\bar\theta^{\eta}  <  8.8 \cdot 10^{-10} \,,\quad 
\bar\theta^{\eta'} <  9.4 \cdot 10^{-10} 
\end{align} 
results for the average ratio $R = m_u/m_d = 0.468$ 
of quark masses calculated in lattice QCD at a scale of  2 GeV~\cite{PDG}. 
Compared to the bound on $\bar\theta$ directly obtained from the experimental constraint on nEDM, 
$\bar\theta < 6\cdot10^{-10}$~\cite{Crewther:1979pi,Pospelov:1999ha,Balla:1999vx}, 
our calculation shows ( this finding is independent of the assumption that CPV $\eta,\eta'$-decays are generated by the same mechanism as the nEDM) 
the very tight experimental limits on the nEDM exclude large contributions 
to $\eta(\eta')\to\pi\pi$ decays beyond that captured by the Peccei-Quinn mechanism. 
The main difference with our calculation is 
that in the Peccei-Quinn mechanism the CPV $\eta(\eta')\pi\pi$ couplings are suppressed 
by $m_\pi^2$ in the chiral limit. We opted to relax thus constraint but 
the effect of this assumption 
is marginal. Note that the fact that our two-loop result does not contain chiral 
divergences essentially means that chiral symmetry does not play a role in our scenario, 
consistent with the assumption that the couplings $f_{\eta\pi\pi}$ and $f_{\eta'\pi\pi}$ 
may not be suppressed by the pion mass squared.

In summary, we derive new stringent upper limits on the CPV decays 
$\eta \to \pi\pi$ and $\eta' \to \pi\pi$. 
The presence of an effective CPV $\eta^{(')} \pi\pi$ interaction in the Lagrangian
leads to an induced nEDM at two loop. We explicitly evaluated a full set of two-loop level of 
Feynman diagrams arising at leading chiral order in relativistic ChPT with the pseudoscalar 
pion-nucleon coupling, which are free from divergences. 
The tight experimental bounds on the nEDM lead to upper limits for $|f_{\eta\pi\pi}|$ 
and $|f_{\eta^\prime\pi\pi}|$ which thus cannot exceed few parts times $10^{-11}$. 
These translate into upper limits for the branching ratios $\mathrm{Br}(\eta \to \pi\pi)$ 
and $\mathrm{Br}(\eta'\to \pi\pi)$, which are  of order $10^{-17}$ or even smaller. 

In the future, we plan 
to continue our study of rare decays of $\eta$ and $\eta'$ mesons. 
In particular, in our scenario only the decays into charged pions are 
strictly speaking constrained. The bound on the neutral decays is obtained 
by isospin symmetry. In presence of isospin symmetry breaking the 
couplings $f_{\eta\pi^+\pi^-}$ and $f_{\eta \to \pi^0\pi^0}$ will be unrelated. 
In this case, with all neutral particles in the loops 
the nEDM can be generated via a magnetic coupling of the photon to the neutron. 

\begin{acknowledgments} 

We are grateful to Vincenzo Cirigliano for helpful discussions.	
A.S.Z. thanks the Mainz Institute for Theoretical Physics (MITP) 
for its hospitality and support of participation in the MITP program 
during which this work was started. 
This work was funded by  
the Carl Zeiss Foundation under Project ``Kepler Center f\"ur Astro- und 
Teilchenphysik: Hochsensitive Nachweistechnik zur Erforschung des
unsichtbaren Universums (Gz: 0653-2.8/581/2)'',
by CONICYT (Chile) PIA/Basal FB0821, by the Russian Federation
program ``Nauka'' (Contract No. 0.1764.GZB.2017), by Tomsk State University
competitiveness improvement program under grant No. 8.1.07.2018, and
by Tomsk Polytechnic University
Competitiveness Enhancement Program (Grant No. VIU-FTI-72/2017) and by DFG (Deutsche Forschungsgesellschaft), project "SFB 1044, Teilprojekt S2".
	
\end{acknowledgments} 

\begin{widetext}
\vspace*{.25cm}
\appendix
\section{Calculation technique of the two-loop integrals}
\label{appendixA}

\begin{figure}[htb]		
	\includegraphics[width=0.45\linewidth, trim={2cm 23cm 7cm 1cm},
	clip]{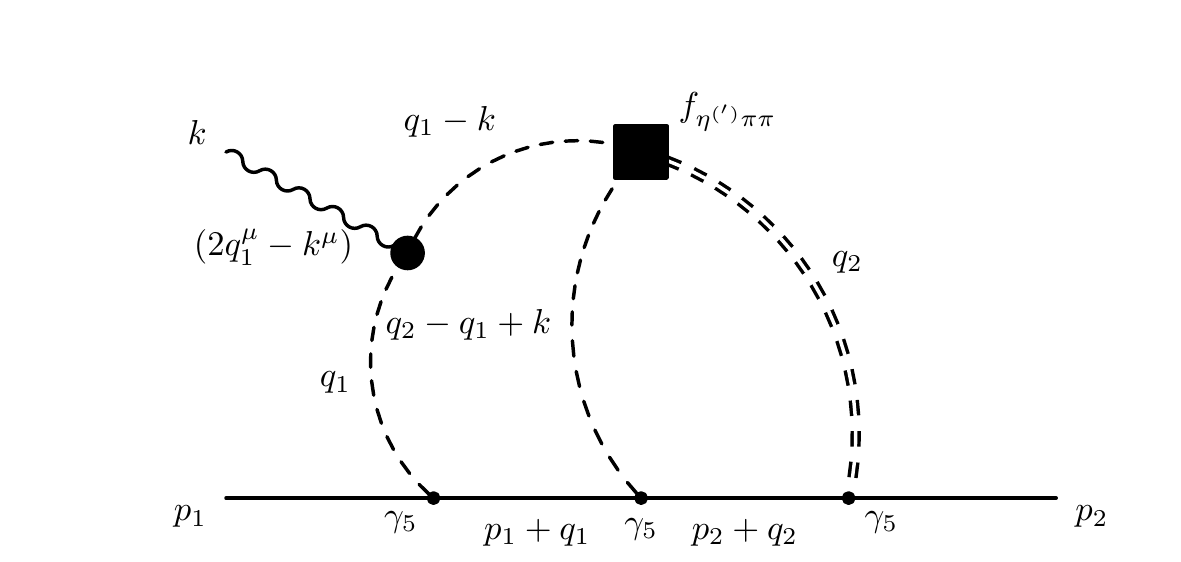}
	\includegraphics[width=0.45\linewidth, trim={2cm 23cm 7cm 1cm},
	clip]{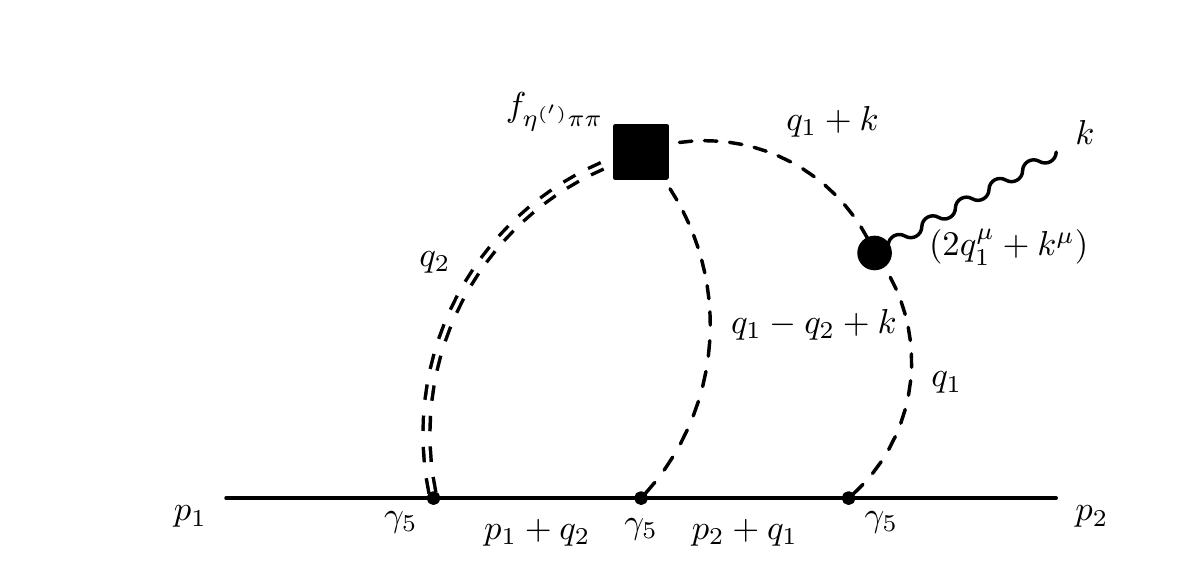}
	\caption{Diagrams e) and l) from Fig.~\ref{two_loop}. 
		The double dashed lines correspond to the $\eta^{(\prime)}$ mesons, 
		the dashed lines to pions and the solid lines to baryon propagators. 
		The CPV $\eta^{(\prime)}\pi\pi$ transition is denoted by the black box.}
	\label{two_loop_2}
\end{figure}

Here we discuss the calculational technique of the two-loop integrals occuring 
in the evaluation of the nEDM (see diagrams in Fig.~\ref{two_loop}). 
Analytic manipulation of the  integrals is performed using 
the package FORM~\cite{Vermaseren:2000nd}. 
For convenience we evaluate the two-loop integrals 
in $d$-dimensions and finally put $d=4$. 

We demonstrate all steps of the calculations for the diagrams 
of Fig.~\ref{two_loop}e and~\ref{two_loop}l, which are shown in more detail 
in Fig.~\ref{two_loop_2}. In particular, the matrix element 
generated by the diagram in Fig.~\ref{two_loop}e is 

\beqn
& &M_{2e}(p_1,p_2) = 
\epsilon_\mu(q) \bar u_N(p_2) \Lambda_{2e}^\mu(p_1,p_2) u_N(p_1)\,, \nonumber\\
& &\Lambda_{2e}^\mu(p_1,p_2) = g_{\rm eff} \,\int \frac{d^dq_1}{(2\pi)^d i} 
                               \int \frac{d^dq_2}{(2\pi)^d i} 
\, \frac{2 q_1^\mu \, i\gamma^5 (m_N + \not\! p_2 + \not\! q_2) 
i\gamma^5 (m_N + \not\! p_1 + \not\! q_1) 
i\gamma^5}{
D_N(p_1+q_1)\,
D_N(p_2+q_2)\, 
D_{\eta^{(')}}(q_2)\,
D_\pi(q_1)\, 
D_\pi(q_1-k)\, 
D_\pi(q_2-q_1+k)}  \,,
\eeqn 
where $u_N(p_1)$ and $\bar u_N(p_2)$ are the nucleon spinors in the initial 
and final state, respectively; $\epsilon_\mu(q)$ is the polarization vector 
of the photon field; $D_H(k) = m_H^2-k^2$ are the scalar denominators  
of virtual particles $H=N,\pi,\eta^{(')}$. 
Here $g_{\rm eff}$ is the effective coupling 
\beqn 
g_{\rm eff} = 4 e g_{\pi NN}^2 g_{\eta^{(')}NN} f_{\eta^{(')}\pi\pi} m_{\eta^{(')}} \,. 
\eeqn 
Using the equation of motion for the nucleon spinors the numerator is reduced to 
$2 i\gamma^5 q_1^\mu \not\! q_2 \not\! q_1$. 
For the string of denominators $D_H$ we apply the Feynman parametrization: 
\beqn 
\frac{1}{D_1\cdots D_6} = \Gamma(6) \, 
\int\limits_0^1 d\alpha_1 \cdots \int\limits_0^1 d\alpha_6 \, 
\frac{\delta\Big(1-\sum\limits_{i=1}^6\alpha_i\Big)}
{\Big(\sum\limits_{i=1}^6 \alpha_i D_i\Big)^6} \,. 
\eeqn 
Then $\Lambda_{2e}^\mu(p_1,p_2)$ takes the form 
\beqn 
\Lambda_{2e}^\mu(p_1,p_2) &=& 2 i\gamma^5 \gamma^\alpha \gamma^\beta 
\, T_{2e}^{\mu\alpha\beta}(p_1,p_2)\,,\nonumber\\
T_{2e}^{\mu\alpha\beta}(p_1,p_2) &=& \Gamma(6) 
\int \frac{d^dq_1}{(2\pi)^d i}
\int \frac{d^dq_2}{(2\pi)^d i} \, k_1^\mu\, k_2^\alpha\, k_1^\beta \, 
\int\limits_0^1 d\alpha_1 \cdots \int\limits_0^1 d\alpha_6 \, 
\delta\Big(1-\sum\limits_{i=1}^6\alpha_i\Big) \,
\frac{1}{(\Delta - q A q - 2 B q)^6}\,.  
\eeqn 
Here 
\beqn 
q A q = \sum\limits_{i,j=1}^2 \, q_i \, A_{ij} \, q_j \,,\quad 
B q   = B_1 q_1 + B_2 q_2 \,, 
\eeqn 
where $A_{ij}$ is the $2\times 2$ matrix 
\beqn 
A_{ij} &=& \left(
\begin{array}{cc}
\alpha_{1456} & - \alpha_6   \\
- \alpha_6    & \alpha_{236} \\
\end{array}
\right)\,, \quad \alpha_{i_1\cdots i_k} = \alpha_{i_1} + \ldots + \alpha_{i_k}\,,
\eeqn 
and
\beqn 
B_1 = p_1 \alpha_1 - k \alpha_{56}\,, \quad B_2 \,=\, p_2 \alpha_2 + k \alpha_6\,, 
\quad \Delta = M_{\eta^{(')}}^2 \alpha_3 + M_\pi^2 \alpha_{456} \,. 
\eeqn
Next, each virtual momentum in the numerator of $T_{2e}^{\mu\alpha\beta}(p_1,p_2)$
can be replaced by the corresponding partial derivative of the 
two-loop integral with respect to $B_1^\alpha$ or $B_2^\alpha$ 
using the substitution 
\beqn 
\frac{q_i^\mu}{(\Delta - q A q - 2 B q)^n} = 
\frac{1}{2 (n-1)} \, \frac{\partial}{\partial B_i^\mu} 
\frac{1}{(\Delta - q A q - 2 B q)^{n-1}} \,. 
\eeqn
In our case we have 
\beqn 
\label{T2e}
T_{2e}^{\mu\alpha\beta}(p_1,p_2) &=& \frac{\Gamma(3)}{2^3} \, 
\frac{\partial^3}{\partial B^\mu_1 \partial B^\beta_1 \partial B^\alpha_2} 
\,  
\int \frac{d^dq_1}{(2\pi)^d i}
\int \frac{d^dq_2}{(2\pi)^d i} 
\int\limits_0^1 d\alpha_1 \cdots \int\limits_0^1 d\alpha_6 \, 
\delta\Big(1-\sum\limits_{i=1}^6\alpha_i\Big) \,
\frac{1}{(\Delta - q A q - 2 B q)^3} \nonumber\\
&=& 
\frac{1}{({\rm det}A)^2} \, \frac{\Gamma(3-d)}{2^3 (4\pi)^d}
\frac{\partial^3}{\partial B^\mu_1 \partial B^\beta_1 \partial B^\alpha_2} 
\Big(\Delta + B A^{-1} B\Big)^{d-3} \,,
\eeqn 
where ${\rm det}A$ and $A^{-1}$ are the determinant and inverse matrix of $A$, 
respectively. Note, at $q^2 = 0$ the term $B A^{-1} B$ is equal to 
\beqn 
B A^{-1} B = m_N^2 \Big( A_{11}^{-1}   \alpha_1^2 
                       + A_{22}^{-1}   \alpha_2^2 
                       + 2 A_{12}^{-1} \alpha_1 \alpha_2\Big) \,.
\eeqn 
Taking the derivatives in Eq.~(\ref{T2e}) we get 
\beqn
\frac{\Gamma(3-d)}{2^3} 
\frac{\partial^3}{\partial B^\mu_1 \partial B^\beta_1 \partial B^\alpha_2}
\Big(\Delta + B A^{-1} B\Big)^{d-3}
&=& \frac{\Gamma(5-d)}{2} \, (\Delta + B A^{-1} B)^{d-5} \, 
\biggl[
g^{\alpha\beta} L_1^\mu    A_{12}^{-1} \, + \, 
g^{\mu\alpha}   L_1^\beta  A_{12}^{-1} \, + \,
g^{\mu\beta}    L_2^\alpha A_{11}^{-1} 
\biggr] 
\nonumber\\
&-& \Gamma(6-d) \, (\Delta + B A^{-1} B)^{d-6} \, L_1^\mu \, L_1^\beta \, L_2^\alpha \,, 
\eeqn 
where $L_1 = B_1 A_{11}^{-1} + B_2 A_{12}^{-1}$ 
and   $L_2 = B_1 A_{12}^{-1} + B_2 A_{22}^{-1}$.

After a straigthforward calculation we keep terms proportional 
to the Dirac structure $i (p_1+p_2)^{\mu} \gamma^5$, 
which due to Gordon identity
\begin{eqnarray} 
i (p_1+p_2)^{\mu} \bar u_N(p_2) \gamma^5 u_N(p_1) = 
\bar u_N(p_2) \sigma^{\mu\nu}k_{\nu}\gamma^5 u_N(p_1)\,,  
\end{eqnarray} 
corresponds to the EDM Dirac structure in Eq.~(\ref{vertex}).  
The contribution of the diagram Fig.~\ref{two_loop}e to the nEDM is  
\beqn 
d_n^{E;e} &=& \frac{g_{\rm eff}}{(4\pi)^4} \, 
\int\limits_0^1 d\alpha_1 \cdots \int\limits_0^1 d\alpha_6 \, 
\frac{\delta\Big(1-\sum\limits_{i=1}^6\alpha_i\Big)}{({\rm det}A)^2 \, 
(\Delta + B A^{-1} B)} \, \biggl[ 
(   -   A_{12}^{-1})^2 (3 \alpha_2 + \alpha_6) 
    +   A_{11}^{-1} A_{22}^{-1} \alpha_6 
    - 3 A_{11}^{-1} A_{12}^{-1} \alpha_1 \nonumber\\
&+& \frac{m_N^2}{\Delta + B A^{-1} B} 
\biggl( (A_{12}^{-1})^3 \alpha_2 (3 \alpha_1\alpha_2 
                      + 2 \alpha_1 \alpha_6
                      + 2 \alpha_2 \alpha_{56}) 
     +   (A_{12}^{-1})^2 A_{22}^{-1} \alpha_2^3 
     + 2 (A_{12}^{-1})^2 A_{11}^{-1} \alpha_1 (\alpha_2 \alpha_{56} 
                                             + \alpha_1 \alpha_6
                                             + 2 \alpha_1 \alpha_2) 
\nonumber\\
     &+& (A_{11}^{-1})^2 A_{12}^{-1} \alpha_1^3 
      +  (A_{11}^{-1})^2 A_{22}^{-1} \alpha_1 (\alpha_1 \alpha_2 
                                           + 2 \alpha_1 \alpha_6 
                                           + 2 \alpha_2 \alpha_{56})  
      -  2 A_{11}^{-1} A_{12}^{-1} A_{22}^{-1} \alpha_2 (\alpha_1 \alpha_6 
                                                    +   \alpha_2 \alpha_{56}) 
\biggr)
\biggr] \,. 
\eeqn 

Using our method we can evaluate the contribution of the other diagrams 
to the nEDM. We only present the final results in terms of integrals over 
Feynman parameters and specify the forms of matrix $A$, vectors $B_1$ and 
$B_2$, and the term $\Delta$.

The contribution of diagrams 2g(h) is given by the expression for 
diagram 2e(l) 
with only one change. The term $\Delta$ must be redefined as 
\beqn 
\Delta = m_\pi^2 \alpha_{345} + m_{\eta^{(')}}^2 \alpha_6 \,. 
\eeqn 

Diagrams 2f(k) give the following contribution 
\beqn 
d_n^{E;f} &=& \frac{4 g_{\rm eff}}{(4\pi)^4} \, 
\int\limits_0^1 d\alpha_1 \cdots \int\limits_0^1 d\alpha_6 \, 
\frac{\delta\Big(1-\sum\limits_{i=1}^6\alpha_i\Big)}{({\rm det}A)^2 \, 
(\Delta + B A^{-1} B)} \, \biggl[ 
( -    A_{12}^{-1})^2 (\alpha_1 - \alpha_2) \nonumber\\
 &+& 3 A_{11}^{-1} A_{12}^{-1} \alpha_1 
  -  3 A_{22}^{-1} A_{12}^{-1} \alpha_2 
+ \frac{m_N^2}{\Delta + B A^{-1} B} 
\biggl( - (A_{12}^{-1})^3 (\alpha_1 - \alpha_2) (2 \alpha_6 \alpha_{12}  
                                             - 3 \alpha_1 \alpha_2) \nonumber\\
&-&  (A_{12}^{-1})^2 A_{22}^{-1} \alpha_2 
     (2 \alpha_6 \alpha_{12} + \alpha_2 (\alpha_2 - 4 \alpha_1)) 
   + (A_{12}^{-1})^2 A_{11}^{-1} \alpha_1 
     (2 \alpha_6 \alpha_{12} + \alpha_1 (\alpha_1 - 4 \alpha_2)) 
\nonumber\\
     &-& (A_{11}^{-1})^2 A_{12}^{-1} \alpha_1^3 
      -  (A_{11}^{-1})^2 A_{22}^{-1} \alpha_1 (2 \alpha_6 \alpha_{12} 
                                             - \alpha_1 \alpha_2)
      +  (A_{22}^{-1})^2 A_{11}^{-1} \alpha_2 (2 \alpha_6 \alpha_{12} 
                                             - \alpha_1 \alpha_2)\nonumber\\
     &+&  (A_{22}^{-1})^2 A_{12}^{-1} \alpha_2^3 
      +   2 A_{11}^{-1} A_{12}^{-1} A_{22}^{-1} \alpha_6 (\alpha_1^2 - \alpha_2^2) 
\biggr)
\biggr]
\eeqn 

Diagrams 2a(b) and 2c(d) result in the expresion 
\beqn 
d_n^{E;a(c)} &=& \frac{6 g_{\rm eff}}{(4\pi)^4} \, 
\int\limits_0^1 d\alpha_1 \cdots \int\limits_0^1 d\alpha_6 \, 
\frac{\delta\Big(1-\sum\limits_{i=1}^6\alpha_i\Big)}{({\rm det}A)^2 \, 
(\Delta_{a(c)} + B A^{-1} B)} \, \biggl[ 
  2 A_{12}^{-1} 
- 6 (A_{12}^{-1})^2 \alpha_2  
- 6 A_{11}^{-1} A_{12}^{-1} \alpha_1 
\nonumber\\
&+& \frac{m_N^2}{\Delta_{a(c)} + B A^{-1} B} 
\biggl(- 4 (A_{12}^{-1})^2 \alpha_2 \alpha_{14}
       - 4  A_{11}^{-1} A_{12}^{-1} \alpha_1 \alpha_{14} 
       - 4  A_{11}^{-1} A_{22}^{-1} \alpha_1 \alpha_2\nonumber\\ 
      &-& 4 A_{12}^{-1} A_{22}^{-1} \alpha_2^2 
       + 2 (A_{12}^{-1})^3 \alpha_2^2 (2 \alpha_4 
                                     + \alpha_1)
       + 2 (A_{12}^{-1})^2 A_{11}^{-1} \alpha_2 \alpha_{14}^2  
       + 2 (A_{12}^{-1})^2 A_{22}^{-1} \alpha_2^3 
\nonumber\\
     &+& 2 (A_{11}^{-1})^2 A_{12}^{-1} \alpha_1 \alpha_{14}^2
      +  2 (A_{11}^{-1})^2 A_{22}^{-1} \alpha_2 (\alpha_1^2 - \alpha_4^2)  
      +  4 A_{11}^{-1} A_{12}^{-1} A_{22}^{-1} \alpha_1 \alpha_2^2 
\biggr)
\biggr]
\eeqn 
Here matrix $A$ is the same as for diagram 2e(l), the 
vectors $B_i$ are 
$B_1 = p_1 \alpha_1 + p_2 \alpha_4$ and $B_2 = p_1 \alpha_2$.  
The $\Delta$ terms are specified as: 
\beqn 
\Delta_a = m_\pi^2 \alpha_{35} + m_{\eta^{(')}}^2 \alpha_6\,, \quad 
\Delta_c = m_\pi^2 \alpha_{56} + m_{\eta^{(')}}^2 \alpha_3\,. 
\eeqn 
One can see that the contributions of diagrams Fig.2a(b) and 2c(d) are 
degenerate in the limit $M_{\eta^{(')}} = M_\pi$. The numerical values for 
these two types of diagrams  at physical values of $\pi$ and $\eta^{(')}$ 
masses are also close to each other.  
  
After restoring the omitted isospin factors and couplings, 
we obtain the total contribution to the nEDM:
\begin{align}\label{finalcoeffs}
d_N^E =& 2 (d_N^{E;a}+d_N^{E;c}+d_N^{E;f}+d_N^{E;g}+d_N^{E;e})
\end{align}

\end{widetext}


\begin{thebibliography}{27}%
	\makeatletter
	\providecommand \@ifxundefined [1]{%
		\@ifx{#1\undefined}
	}%
	\providecommand \@ifnum [1]{%
		\ifnum #1\expandafter \@firstoftwo
		\else \expandafter \@secondoftwo
		\fi
	}%
	\providecommand \@ifx [1]{%
		\ifx #1\expandafter \@firstoftwo
		\else \expandafter \@secondoftwo
		\fi
	}%
	\providecommand \natexlab [1]{#1}%
	\providecommand \enquote  [1]{``#1''}%
	\providecommand \bibnamefont  [1]{#1}%
	\providecommand \bibfnamefont [1]{#1}%
	\providecommand \citenamefont [1]{#1}%
	\providecommand \href@noop [0]{\@secondoftwo}%
	\providecommand \href [0]{\begingroup \@sanitize@url \@href}%
	\providecommand \@href[1]{\@@startlink{#1}\@@href}%
	\providecommand \@@href[1]{\endgroup#1\@@endlink}%
	\providecommand \@sanitize@url [0]{\catcode `\\12\catcode `\$12\catcode
		`\&12\catcode `\#12\catcode `\^12\catcode `\_12\catcode `\%12\relax}%
	\providecommand \@@startlink[1]{}%
	\providecommand \@@endlink[0]{}%
	\providecommand \url  [0]{\begingroup\@sanitize@url \@url }%
	\providecommand \@url [1]{\endgroup\@href {#1}{\urlprefix }}%
	\providecommand \urlprefix  [0]{URL }%
	\providecommand \Eprint [0]{\href }%
	\providecommand \doibase [0]{http://dx.doi.org/}%
	\providecommand \selectlanguage [0]{\@gobble}%
	\providecommand \bibinfo  [0]{\@secondoftwo}%
	\providecommand \bibfield  [0]{\@secondoftwo}%
	\providecommand \translation [1]{[#1]}%
	\providecommand \BibitemOpen [0]{}%
	\providecommand \bibitemStop [0]{}%
	\providecommand \bibitemNoStop [0]{.\EOS\space}%
	\providecommand \EOS [0]{\spacefactor3000\relax}%
	\providecommand \BibitemShut  [1]{\csname bibitem#1\endcsname}%
	\let\auto@bib@innerbib\@empty
	\bibitem [{\citenamefont {Sakharov}(1967)}]{Sakharov:1967dj}%
	\BibitemOpen
	\bibfield  {author} {\bibinfo {author} {\bibfnamefont {A.~D.}\ \bibnamefont
			{Sakharov}},\ }\href {\doibase 10.1070/PU1991v034n05ABEH002497} {\bibfield
		{journal} {\bibinfo  {journal} {Pisma Zh. Eksp. Teor. Fiz.}\ }\textbf
		{\bibinfo {volume} {5}},\ \bibinfo {pages} {32} (\bibinfo {year} {1967})},\
	\bibinfo {note} {[Usp. Fiz. Nauk 161, no. 5, 61 (1991)]}\BibitemShut
	{NoStop}%
	\bibitem [{\citenamefont {Chupp}\ \emph {et~al.}(2017)\citenamefont {Chupp},
		\citenamefont {Fierlinger}, \citenamefont {Ramsey-Musolf},\ and\
		\citenamefont {Singh}}]{Chupp:2017rkp}%
	\BibitemOpen
	\bibfield  {author} {\bibinfo {author} {\bibfnamefont {T.}~\bibnamefont
			{Chupp}}, \bibinfo {author} {\bibfnamefont {P.}~\bibnamefont {Fierlinger}},
		\bibinfo {author} {\bibfnamefont {M.}~\bibnamefont {Ramsey-Musolf}}, \ and\
		\bibinfo {author} {\bibfnamefont {J.}~\bibnamefont {Singh}},\ }\href@noop {}
	{\  (\bibinfo {year} {2017})},\ \Eprint {http://arxiv.org/abs/1710.02504}
	{arXiv:1710.02504} \BibitemShut {NoStop}%
	\bibitem [{\citenamefont {Aaij}\ \emph {et~al.}(2017)\citenamefont {Aaij} \emph
		{et~al.}}]{Aaij:2016jaa}%
	\BibitemOpen
	\bibfield  {author} {\bibinfo {author} {\bibfnamefont {R.}~\bibnamefont
			{Aaij}} \emph {et~al.} (\bibinfo {collaboration} {LHCb Collaboration}),\
	}\href {\doibase 10.1016/j.physletb.2016.11.032} {\bibfield  {journal}
		{\bibinfo  {journal} {Phys. Lett.}\ }\textbf {\bibinfo {volume} {B764}},\
		\bibinfo {pages} {233} (\bibinfo {year} {2017})}\BibitemShut {NoStop}%
	\bibitem [{\citenamefont {Pitschmann}\ \emph {et~al.}(2015)\citenamefont
		{Pitschmann}, \citenamefont {Seng}, \citenamefont {Roberts},\ and\
		\citenamefont {Schmidt}}]{Seng:2014lea}%
	\BibitemOpen
	\bibfield  {author} {\bibinfo {author} {\bibfnamefont {M.}~\bibnamefont
			{Pitschmann}}, \bibinfo {author} {\bibfnamefont {C.-Y.}\ \bibnamefont
			{Seng}}, \bibinfo {author} {\bibfnamefont {C.~D.}\ \bibnamefont {Roberts}}, \
		and\ \bibinfo {author} {\bibfnamefont {S.~M.}\ \bibnamefont {Schmidt}},\
	}\href@noop {} {\bibfield  {journal} {\bibinfo  {journal} {Phys. Rev.}\
		}\textbf {\bibinfo {volume} {D91}},\ \bibinfo {pages} {074004} (\bibinfo
		{year} {2015})}\BibitemShut {NoStop}%
	\bibitem [{\citenamefont {Pendlebury}\ \emph {et~al.}(2015)\citenamefont
		{Pendlebury} \emph {et~al.}}]{Afach:2015sja}%
	\BibitemOpen
	\bibfield  {author} {\bibinfo {author} {\bibfnamefont {J.~M.}\ \bibnamefont
			{Pendlebury}} \emph {et~al.},\ }\href {\doibase 10.1103/PhysRevD.92.092003}
	{\bibfield  {journal} {\bibinfo  {journal} {Phys. Rev.}\ }\textbf {\bibinfo
			{volume} {D92}},\ \bibinfo {pages} {092003} (\bibinfo {year} {2015})},\
	\Eprint {http://arxiv.org/abs/1509.04411} {arXiv:1509.04411 [hep-ex]}
	\BibitemShut {NoStop}%
	\bibitem [{\citenamefont {Tanabashi}\ \emph {et~al.}(2018)\citenamefont
		{Tanabashi} \emph {et~al.}}]{PDG}%
	\BibitemOpen
	\bibfield  {author} {\bibinfo {author} {\bibfnamefont {M.}~\bibnamefont
			{Tanabashi}} \emph {et~al.} (\bibinfo {collaboration} {Particle Data
			Group}),\ }\href {\doibase 10.1103/PhysRevD.98.030001} {\bibfield  {journal}
		{\bibinfo  {journal} {Phys. Rev.}\ }\textbf {\bibinfo {volume} {D98}},\
		\bibinfo {pages} {030001} (\bibinfo {year} {2018})}\BibitemShut {NoStop}%
	\bibitem [{\citenamefont {Diakonov}\ and\ \citenamefont
		{Eides}(1981)}]{Diakonov:1981nv}%
	\BibitemOpen
	\bibfield  {author} {\bibinfo {author} {\bibfnamefont {D.}~\bibnamefont
			{Diakonov}}\ and\ \bibinfo {author} {\bibfnamefont {M.~I.}\ \bibnamefont
			{Eides}},\ }\href@noop {} {\bibfield  {journal} {\bibinfo  {journal} {Sov.
				Phys. JETP}\ }\textbf {\bibinfo {volume} {54}},\ \bibinfo {pages} {232}
		(\bibinfo {year} {1981})},\ \bibinfo {note} {[Zh. Eksp. Teor.
		Fiz.81,434(1981)]}\BibitemShut {NoStop}%
	\bibitem [{\citenamefont {Witten}(1979)}]{Witten:1979vv}%
	\BibitemOpen
	\bibfield  {author} {\bibinfo {author} {\bibfnamefont {E.}~\bibnamefont
			{Witten}},\ }\href {\doibase 10.1016/0550-3213(79)90031-2} {\bibfield
		{journal} {\bibinfo  {journal} {Nucl. Phys.}\ }\textbf {\bibinfo {volume}
			{B156}},\ \bibinfo {pages} {269} (\bibinfo {year} {1979})}\BibitemShut
	{NoStop}%
	\bibitem [{\citenamefont {Peccei}\ and\ \citenamefont
		{Quinn}(1977)}]{Peccei:1977hh}%
	\BibitemOpen
	\bibfield  {author} {\bibinfo {author} {\bibfnamefont {R.~D.}\ \bibnamefont
			{Peccei}}\ and\ \bibinfo {author} {\bibfnamefont {H.~R.}\ \bibnamefont
			{Quinn}},\ }\href {\doibase 10.1103/PhysRevLett.38.1440} {\bibfield
		{journal} {\bibinfo  {journal} {Phys. Rev. Lett.}\ }\textbf {\bibinfo
			{volume} {38}},\ \bibinfo {pages} {1440} (\bibinfo {year}
		{1977})}\BibitemShut {NoStop}%
	\bibitem [{\citenamefont {Castillo-Felisola}\ \emph {et~al.}(2015)\citenamefont
		{Castillo-Felisola}, \citenamefont {Corral}, \citenamefont {Kovalenko},
		\citenamefont {Schmidt},\ and\ \citenamefont
		{Lyubovitskij}}]{Castillo-Felisola:2015ema}%
	\BibitemOpen
	\bibfield  {author} {\bibinfo {author} {\bibfnamefont {O.}~\bibnamefont
			{Castillo-Felisola}}, \bibinfo {author} {\bibfnamefont {C.}~\bibnamefont
			{Corral}}, \bibinfo {author} {\bibfnamefont {S.}~\bibnamefont {Kovalenko}},
		\bibinfo {author} {\bibfnamefont {I.}~\bibnamefont {Schmidt}}, \ and\
		\bibinfo {author} {\bibfnamefont {V.~E.}\ \bibnamefont {Lyubovitskij}},\
	}\href {\doibase 10.1103/PhysRevD.91.085017} {\bibfield  {journal} {\bibinfo
			{journal} {Phys. Rev.}\ }\textbf {\bibinfo {volume} {D91}},\ \bibinfo {pages}
		{085017} (\bibinfo {year} {2015})},\ \Eprint
	{http://arxiv.org/abs/1502.03694} {arXiv:1502.03694 [hep-ph]} \BibitemShut
	{NoStop}%
	\bibitem [{\citenamefont {Crewther}\ \emph {et~al.}(1979)\citenamefont
		{Crewther}, \citenamefont {Di~Vecchia}, \citenamefont {Veneziano},\ and\
		\citenamefont {Witten}}]{Crewther:1979pi}%
	\BibitemOpen
	\bibfield  {author} {\bibinfo {author} {\bibfnamefont {R.~J.}\ \bibnamefont
			{Crewther}}, \bibinfo {author} {\bibfnamefont {P.}~\bibnamefont
			{Di~Vecchia}}, \bibinfo {author} {\bibfnamefont {G.}~\bibnamefont
			{Veneziano}}, \ and\ \bibinfo {author} {\bibfnamefont {E.}~\bibnamefont
			{Witten}},\ }\href@noop {} {\bibfield  {journal} {\bibinfo  {journal} {Phys.
				Lett.}\ }\textbf {\bibinfo {volume} {88B}},\ \bibinfo {pages} {123} (\bibinfo
		{year} {1979})}\BibitemShut {NoStop}%
	\bibitem [{\citenamefont {Al~Ghoul}\ \emph {et~al.}(2016)\citenamefont
		{Al~Ghoul} \emph {et~al.}}]{Ghoul:2015ifw}%
	\BibitemOpen
	\bibfield  {author} {\bibinfo {author} {\bibfnamefont {H.}~\bibnamefont
			{Al~Ghoul}} \emph {et~al.} (\bibinfo {collaboration} {GlueX Collaboration}),\
	}\bibfield  {booktitle} {\emph {\bibinfo {booktitle} {{Proceedings, 16th
					International Conference on Hadron Spectroscopy (Hadron 2015): Newport News,
					Virginia, USA, September 13-18, 2015}}},\ }\href {\doibase 10.1063/1.4949369}
	{\bibfield  {journal} {\bibinfo  {journal} {AIP Conf. Proc.}\ }\textbf
		{\bibinfo {volume} {1735}},\ \bibinfo {pages} {020001} (\bibinfo {year}
		{2016})},\ \Eprint {http://arxiv.org/abs/1512.03699} {arXiv:1512.03699
		[nucl-ex]} \BibitemShut {NoStop}%
	\bibitem [{\citenamefont {Gorchtein}(2008)}]{Gorchtein:2008pe}%
	\BibitemOpen
	\bibfield  {author} {\bibinfo {author} {\bibfnamefont {M.}~\bibnamefont
			{Gorchtein}},\ }\href@noop {} {\  (\bibinfo {year} {2008})},\ \Eprint
	{http://arxiv.org/abs/0803.2906} {arXiv:0803.2906 [hep-ph]} \BibitemShut
	{NoStop}%
	\bibitem [{\citenamefont {Gutsche}\ \emph {et~al.}(2017)\citenamefont
		{Gutsche}, \citenamefont {Hiller~Blin}, \citenamefont {Kovalenko},
		\citenamefont {Kuleshov}, \citenamefont {Lyubovitskij}, \citenamefont
		{Vicente~Vacas},\ and\ \citenamefont {Zhevlakov}}]{Gutsche:2016jap}%
	\BibitemOpen
	\bibfield  {author} {\bibinfo {author} {\bibfnamefont {T.}~\bibnamefont
			{Gutsche}}, \bibinfo {author} {\bibfnamefont {A.~N.}\ \bibnamefont
			{Hiller~Blin}}, \bibinfo {author} {\bibfnamefont {S.}~\bibnamefont
			{Kovalenko}}, \bibinfo {author} {\bibfnamefont {S.}~\bibnamefont {Kuleshov}},
		\bibinfo {author} {\bibfnamefont {V.~E.}\ \bibnamefont {Lyubovitskij}},
		\bibinfo {author} {\bibfnamefont {M.~J.}\ \bibnamefont {Vicente~Vacas}}, \
		and\ \bibinfo {author} {\bibfnamefont {A.}~\bibnamefont {Zhevlakov}},\ }\href
	{\doibase 10.1103/PhysRevD.95.036022} {\bibfield  {journal} {\bibinfo
			{journal} {Phys. Rev.}\ }\textbf {\bibinfo {volume} {D95}},\ \bibinfo {pages}
		{036022} (\bibinfo {year} {2017})},\ \Eprint
	{http://arxiv.org/abs/1612.02276} {arXiv:1612.02276 [hep-ph]} \BibitemShut
	{NoStop}%
	\bibitem [{\citenamefont {Pich}\ and\ \citenamefont
		{de~Rafael}(1991)}]{Pich:1991fq}%
	\BibitemOpen
	\bibfield  {author} {\bibinfo {author} {\bibfnamefont {A.}~\bibnamefont
			{Pich}}\ and\ \bibinfo {author} {\bibfnamefont {E.}~\bibnamefont
			{de~Rafael}},\ }\href {\doibase {10.1016/0550-3213(91)90019-T}} {\bibfield
		{journal} {\bibinfo  {journal} {Nucl. Phys.}\ }\textbf {\bibinfo {volume}
			{B367}},\ \bibinfo {pages} {313} (\bibinfo {year} {1991})}\BibitemShut
	{NoStop}%
	\bibitem [{\citenamefont {Shifman}\ \emph {et~al.}(1980)\citenamefont
		{Shifman}, \citenamefont {Vainshtein},\ and\ \citenamefont
		{Zakharov}}]{Shifman:1979if}%
	\BibitemOpen
	\bibfield  {author} {\bibinfo {author} {\bibfnamefont {M.~A.}\ \bibnamefont
			{Shifman}}, \bibinfo {author} {\bibfnamefont {A.~I.}\ \bibnamefont
			{Vainshtein}}, \ and\ \bibinfo {author} {\bibfnamefont {V.~I.}\ \bibnamefont
			{Zakharov}},\ }\href {\doibase 10.1016/0550-3213(80)90209-6} {\bibfield
		{journal} {\bibinfo  {journal} {Nucl. Phys.}\ }\textbf {\bibinfo {volume}
			{B166}},\ \bibinfo {pages} {493} (\bibinfo {year} {1980})}\BibitemShut
	{NoStop}%
	\bibitem [{\citenamefont {Harris}\ \emph {et~al.}(1999)\citenamefont {Harris}
		\emph {et~al.}}]{Harris:1999jx}%
	\BibitemOpen
	\bibfield  {author} {\bibinfo {author} {\bibfnamefont {P.~G.}\ \bibnamefont
			{Harris}} \emph {et~al.},\ }\href {\doibase 10.1103/PhysRevLett.82.904}
	{\bibfield  {journal} {\bibinfo  {journal} {Phys. Rev. Lett.}\ }\textbf
		{\bibinfo {volume} {82}},\ \bibinfo {pages} {904} (\bibinfo {year}
		{1999})}\BibitemShut {NoStop}%
	\bibitem [{\citenamefont {Baker}\ \emph {et~al.}(2006)\citenamefont {Baker}
		\emph {et~al.}}]{Baker:2006ts}%
	\BibitemOpen
	\bibfield  {author} {\bibinfo {author} {\bibfnamefont {C.~A.}\ \bibnamefont
			{Baker}} \emph {et~al.},\ }\href {\doibase 10.1103/PhysRevLett.97.131801}
	{\bibfield  {journal} {\bibinfo  {journal} {Phys. Rev. Lett.}\ }\textbf
		{\bibinfo {volume} {97}},\ \bibinfo {pages} {131801} (\bibinfo {year}
		{2006})},\ \Eprint {http://arxiv.org/abs/hep-ex/0602020}
	{arXiv:hep-ex/0602020 [hep-ex]} \BibitemShut {NoStop}%
	\bibitem [{\citenamefont {Pospelov}\ and\ \citenamefont
		{Ritz}(1999)}]{Pospelov:1999ha}%
	\BibitemOpen
	\bibfield  {author} {\bibinfo {author} {\bibfnamefont {M.}~\bibnamefont
			{Pospelov}}\ and\ \bibinfo {author} {\bibfnamefont {A.}~\bibnamefont
			{Ritz}},\ }\href {\doibase 10.1103/PhysRevLett.83.2526} {\bibfield  {journal}
		{\bibinfo  {journal} {Phys. Rev. Lett.}\ }\textbf {\bibinfo {volume} {83}},\
		\bibinfo {pages} {2526} (\bibinfo {year} {1999})},\ \Eprint
	{http://arxiv.org/abs/hep-ph/9904483} {arXiv:hep-ph/9904483 [hep-ph]}
	\BibitemShut {NoStop}%
	\bibitem [{\citenamefont {Balla}\ \emph {et~al.}(1999)\citenamefont {Balla},
		\citenamefont {Blotz},\ and\ \citenamefont {Goeke}}]{Balla:1999vx}%
	\BibitemOpen
	\bibfield  {author} {\bibinfo {author} {\bibfnamefont {J.}~\bibnamefont
			{Balla}}, \bibinfo {author} {\bibfnamefont {A.}~\bibnamefont {Blotz}}, \ and\
		\bibinfo {author} {\bibfnamefont {K.}~\bibnamefont {Goeke}},\ }\href
	{\doibase 10.1103/PhysRevD.59.056005} {\bibfield  {journal} {\bibinfo
			{journal} {Phys. Rev.}\ }\textbf {\bibinfo {volume} {D59}},\ \bibinfo {pages}
		{056005} (\bibinfo {year} {1999})}\BibitemShut {NoStop}%
	\bibitem [{\citenamefont {Lyubovitskij}\ \emph {et~al.}(2001)\citenamefont
		{Lyubovitskij}, \citenamefont {Gutsche}, \citenamefont {Faessler},\ and\
		\citenamefont {Vinh~Mau}}]{Lyubovitskij:2001fv}%
	\BibitemOpen
	\bibfield  {author} {\bibinfo {author} {\bibfnamefont {V.~E.}\ \bibnamefont
			{Lyubovitskij}}, \bibinfo {author} {\bibfnamefont {T.}~\bibnamefont
			{Gutsche}}, \bibinfo {author} {\bibfnamefont {A.}~\bibnamefont {Faessler}}, \
		and\ \bibinfo {author} {\bibfnamefont {R.}~\bibnamefont {Vinh~Mau}},\ }\href
	{\doibase 10.1016/S0370-2693(01)01142-X} {\bibfield  {journal} {\bibinfo
			{journal} {Phys. Lett.}\ }\textbf {\bibinfo {volume} {B520}},\ \bibinfo
		{pages} {204} (\bibinfo {year} {2001})},\ \Eprint
	{http://arxiv.org/abs/hep-ph/0108134} {arXiv:hep-ph/0108134 [hep-ph]}
	\BibitemShut {NoStop}%
	\bibitem [{\citenamefont {Lyubovitskij}\ \emph {et~al.}(2002)\citenamefont
		{Lyubovitskij}, \citenamefont {Gutsche}, \citenamefont {Faessler},\ and\
		\citenamefont {Vinh~Mau}}]{Lyubovitskij:2001zn}%
	\BibitemOpen
	\bibfield  {author} {\bibinfo {author} {\bibfnamefont {V.~E.}\ \bibnamefont
			{Lyubovitskij}}, \bibinfo {author} {\bibfnamefont {T.}~\bibnamefont
			{Gutsche}}, \bibinfo {author} {\bibfnamefont {A.}~\bibnamefont {Faessler}}, \
		and\ \bibinfo {author} {\bibfnamefont {R.}~\bibnamefont {Vinh~Mau}},\ }\href
	{\doibase 10.1103/PhysRevC.65.025202} {\bibfield  {journal} {\bibinfo
			{journal} {Phys. Rev.}\ }\textbf {\bibinfo {volume} {C65}},\ \bibinfo {pages}
		{025202} (\bibinfo {year} {2002})},\ \Eprint
	{http://arxiv.org/abs/hep-ph/0109213} {arXiv:hep-ph/0109213 [hep-ph]}
	\BibitemShut {NoStop}%
	\bibitem [{\citenamefont {Lensky}\ and\ \citenamefont
		{Pascalutsa}(2010)}]{Lensky:2009uv}%
	\BibitemOpen
	\bibfield  {author} {\bibinfo {author} {\bibfnamefont {V.}~\bibnamefont
			{Lensky}}\ and\ \bibinfo {author} {\bibfnamefont {V.}~\bibnamefont
			{Pascalutsa}},\ }\href {\doibase 10.1140/epjc/s10052-009-1183-z} {\bibfield
		{journal} {\bibinfo  {journal} {Eur. Phys. J.}\ }\textbf {\bibinfo {volume}
			{C65}},\ \bibinfo {pages} {195} (\bibinfo {year} {2010})},\ \Eprint
	{http://arxiv.org/abs/0907.0451} {arXiv:0907.0451 [hep-ph]} \BibitemShut
	{NoStop}%
	\bibitem [{\citenamefont {Tiator}\ \emph {et~al.}(2018)\citenamefont {Tiator},
		\citenamefont {Gorchteyn}, \citenamefont {Kashevarov}, \citenamefont
		{Nikonov}, \citenamefont {Ostrick}, \citenamefont {Hadzimehmedovic},
		\citenamefont {Omerovic}, \citenamefont {Osmanovic}, \citenamefont {Stahov},\
		and\ \citenamefont {Svarc}}]{Tiator:2018heh}%
	\BibitemOpen
	\bibfield  {author} {\bibinfo {author} {\bibfnamefont {L.}~\bibnamefont
			{Tiator}}, \bibinfo {author} {\bibfnamefont {M.}~\bibnamefont {Gorchteyn}},
		\bibinfo {author} {\bibfnamefont {V.~L.}\ \bibnamefont {Kashevarov}},
		\bibinfo {author} {\bibfnamefont {K.}~\bibnamefont {Nikonov}}, \bibinfo
		{author} {\bibfnamefont {M.}~\bibnamefont {Ostrick}}, \bibinfo {author}
		{\bibfnamefont {M.}~\bibnamefont {Hadzimehmedovic}}, \bibinfo {author}
		{\bibfnamefont {R.}~\bibnamefont {Omerovic}}, \bibinfo {author}
		{\bibfnamefont {H.}~\bibnamefont {Osmanovic}}, \bibinfo {author}
		{\bibfnamefont {J.}~\bibnamefont {Stahov}}, \ and\ \bibinfo {author}
		{\bibfnamefont {A.}~\bibnamefont {Svarc}},\ }\href@noop {} {\  (\bibinfo
		{year} {2018})},\ \Eprint {http://arxiv.org/abs/1807.04525} {arXiv:1807.04525
		[nucl-th]} \BibitemShut {NoStop}%
	\bibitem [{\citenamefont {Guo}\ \emph {et~al.}(2012)\citenamefont {Guo},
		\citenamefont {Kubis},\ and\ \citenamefont {Wirzba}}]{Guo:2011ir}%
	\BibitemOpen
	\bibfield  {author} {\bibinfo {author} {\bibfnamefont {F.-K.}\ \bibnamefont
			{Guo}}, \bibinfo {author} {\bibfnamefont {B.}~\bibnamefont {Kubis}}, \ and\
		\bibinfo {author} {\bibfnamefont {A.}~\bibnamefont {Wirzba}},\ }\href
	{\doibase 10.1103/PhysRevD.85.014014} {\bibfield  {journal} {\bibinfo
			{journal} {Phys. Rev.}\ }\textbf {\bibinfo {volume} {D85}},\ \bibinfo {pages}
		{014014} (\bibinfo {year} {2012})},\ \Eprint {http://arxiv.org/abs/1111.5949}
	{arXiv:1111.5949 [hep-ph]} \BibitemShut {NoStop}%
	\bibitem [{\citenamefont {Gasser}\ and\ \citenamefont
		{Leutwyler}(1982)}]{Gasser:1982ap}%
	\BibitemOpen
	\bibfield  {author} {\bibinfo {author} {\bibfnamefont {J.}~\bibnamefont
			{Gasser}}\ and\ \bibinfo {author} {\bibfnamefont {H.}~\bibnamefont
			{Leutwyler}},\ }\href {\doibase 10.1016/0370-1573(82)90035-7} {\bibfield
		{journal} {\bibinfo  {journal} {Phys. Rept.}\ }\textbf {\bibinfo {volume}
			{87}},\ \bibinfo {pages} {77} (\bibinfo {year} {1982})}\BibitemShut {NoStop}%
	\bibitem [{\citenamefont {Vermaseren}(2000)}]{Vermaseren:2000nd}%
	\BibitemOpen
	\bibfield  {author} {\bibinfo {author} {\bibfnamefont {J.~A.~M.}\ 
			\bibnamefont {Vermaseren}},\ }\href@noop {} {\  (\bibinfo {year} {2000})},\
	\Eprint {http://arxiv.org/abs/math-ph/0010025} {arXiv:math-ph/0010025
		[math-ph]} \BibitemShut {NoStop}%
\end{thebibliography}
%
\end{document}